\documentclass[sigconf]{acmart}

\usepackage{booktabs} 
\usepackage{ccicons}  
\usepackage{todonotes}
\usepackage{subfig}
\graphicspath{{figures/}}
\usepackage[english]{babel}
\usepackage[autostyle=true,english=american]{csquotes}
\usepackage{gensymb}
\usepackage{balance}
\usepackage[most]{tcolorbox}
\usepackage{enumitem}
\setlist{leftmargin=5mm, noitemsep, topsep=0pt}
\usepackage{balance}
\usepackage{setspace}

\usepackage{acronym}
\newacro{AI}[AI]{Artificial Intelligence}
\newacro{UI}[UI]{user interface}
\newacro{GUI}[GUI]{graphical user interface}
\newacro{TLX}[TLX]{NASA-Task Load Index}
\newacro{RTLX}[Raw-TLX]{NASA Raw-Task Load Index}
\newacro{ER}[ER]{error rate}
\newacro{TCT}[TCT]{task completion time}
\newacro{HCI}[HCI]{Human-Computer Interaction}
\newacro{UX}[UX]{user experience}
\newacro{HFE}[HFE]{Human Factors and Ergonomics}
\newacro{cuDNN}[cuDNN]{CUDA Deep Neural Network library}
\newacro{RMSE}[RMSE]{root mean squared error}
\newacro{HMD}[HMD]{Head-Mounted Display}
\newacro{RF}[RF]{Random Forest}
\newacro{GP}[GP]{Gaussian process, long-plural = Gaussian processes}
\newacro{KNN}[\textit{k}NN]{\textit{k}-nearest neighbor}
\newacro{NN}[NN]{Neural Network}
\newacro{DNN}[DNN]{ Deep Neural Network}
\newacro{CNN}[CNN]{Convolutional Neural Network}
\newacro{FCL}[FCL]{fully connected layer}
\newacro{BoD}[BoD]{Back-of-Device}
\newacro{FOV}[FoV]{field of view}
\newacro{RW}[RW]{Real World}
\newacro{IFRC}[IFRC]{index finger ray cast}
\newacro{FRC}[FRC]{forearm ray cast}
\newacro{EFRC}[EFRC]{eye-finger ray cast}
\newacro{HRC}[HRC]{Human-Robot Collaboration}
\newacro{HRI}[HRI]{Human-Robot Interaction}
\newacro{6DOF}[6DOF]{six-degree-of-freedom}
\newacro{LOOCV}[LOOCV]{leave-one-out cross-validation}
\newacro{CV}[CV]{cross-validation}
\newacro{RM}[RM]{repeated measure}
\newacro{ANOVA}[ANOVA]{analysis of variance}
\newacro{RMANOVA}[RM-ANOVA]{repeated measures analysis of variance}
\newacro{AGATe}[AGATe]{AGreement Analysis Toolkit}
\newacro{GHoST}[GHoST]{Gesture Heatmap Toolkit Gesture Heatmaps Toolkit}
\newacro{GREAT}[GREAT]{Gesture Relative Accuracy Toolkit}
\newacro{GRT}[GRT]{Gesture Recognition Toolkit}
\newacro{DTW}[DTW]{Dynamic Time Warping}
\newacro{LHRD}[LHRD]{large high resolution display}
\newacro{GEQ}[GEQ]{Game Experience Questionnaire}
\newacro{SPGQ}[SPGQ]{Social Presence Gaming Questionnaire}
\newacro{JND}[JND]{just-noticeable difference}
\newacro{SUS}[SUS]{system usability scale}
\newacro{CSCW}[CSCW]{computer-supported cooperative work}
\newacro{CAD}[CAD]{computer-aided design}
\newacro{MR}[MR]{Mixed Reality}
\newacro{CVE}[CVE]{Collaborative Virtual Environment}
\newacro{AR}[AR]{Augmented Reality}
\newacro{AV}[AV]{Augmented Virtuality}
\newacro{VR}[VR]{Virtual Reality}
\newacro{PRISMA}[PRISMA]{Preferred Reporting Items for Systematic Reviews}
\newacro{PRISMA-Scope}[PRISMA-ScR]{Meta-Analyses Extension for Scoping Reviews}
\newacro{TF-IDF}[TF-IDF]{Term Frequency-Inverse Document Frequency}
\newacro{TF}[TF]{Term Frequency}
\newacro{AVs}[AVs]{Automated Vehicles}
\newacro{eHMIs}[eHMIs]{external Human-machine interfaces}
\newacro{SAR}[SAR]{Spatial Augmented Reality}
\newacro{IFR}[IFR]{International Federation of Robotics}
\newacro{ADLs}[ADLs]{Activities of Daily Living}
\newacro{LED}[LED]{Light-Emitting Diode}
\newacro{DoF}[DoF]{degrees-of-freedom} \newacroplural{DoF}[DoFs]{degrees-of-freedom}
\newacro{HHC}[HHC]{Human-Human Collaboration}
\newacro{IDF}[IDF]{Inverse Document Frequency}
\newacro{PAR}[PAR]{physically assistive robot}
\newacro{CBPR}[CBPR]{Community Based Participatory Research}
\newacro{ADL}[ADL]{activity of daily living} \newacroplural{ADL}[ADLs]{activities of daily living}

\AtBeginDocument{%
  \providecommand\BibTeX{{%
    \normalfont B\kern-0.5em{\scshape i\kern-0.25em b}\kern-0.8em\TeX}}}

\setcopyright{rightsretained}
\acmDOI{}

\acmConference[A3DE @ HRI'24]{Assistive Applications, Accessibility, and Disability Ethics (A3DE @ HRI)}{Mar 15, 2024}{Boulder, CO}
\acmBooktitle{Assistive Applications, Accessibility, and Disability Ethics (A3DE @ HRI)} 
\acmPrice{}
\acmISBN{}

\begin{document}

\title[Multiple Ways of Working with Users to Develop Physically Assistive Robots]{Multiple Ways of Working with Users\\ to Develop Physically Assistive Robots}

\author{Amal Nanavati}
\orcid{0000-0001-5380-7834}
\email{amaln@cs.washington.edu}
\affiliation{
    \institution{University of Washington}
    \city{Seattle}
    \country{United States of America}
}
\authornotemark[1]
\author{Max Pascher}
\orcid{0000-0002-6847-0696}
\email{max.pascher@udo.edu}
\affiliation{
    \institution{TU Dortmund University}
    \city{Dortmund}
    \country{Germany}
}
\affiliation{
    \institution{University of Duisburg-Essen}
    \city{Essen}
    \country{Germany}
}
\authornotemark[1]

\author{Vinitha Ranganeni}
\orcid{0000-0001-7789-7221}
\author{Ethan K. Gordon}
\orcid{0000-0003-1621-2342}
\author{Taylor Kessler Faulkner}
\orcid{0000-0002-5838-0021}
\email{{vinitha, ekgordon, taylorkf}@cs.washington.edu}
\affiliation{
    \institution{University of Washington}
    \city{Seattle}
    \country{United States of America}
}

\author{Siddhartha S. Srinivasa}
\orcid{0000-0002-5091-106X}
\author{Maya Cakmak}
\orcid{0000-0001-8457-6610}
\email{{siddh, mcakmak}@cs.washington.edu}
\affiliation{
    \institution{University of Washington}
    \city{Seattle}
    \country{United States of America}
}

\author{Patrícia Alves-Oliveira}
\orcid{0000-0002-0133-2432}
\email{robopati@umich.edu}
\affiliation{
    \institution{University of Michigan}
    \city{Ann Arbor}
    \country{United States of America}
}

\author{Jens Gerken}
\orcid{0000-0002-0634-3931}
\email{jens.gerken@udo.edu}
\affiliation{
    \institution{TU Dortmund University}
    \city{Dortmund}
    \country{Germany}
}

\renewcommand{\shortauthors}{A. Nanavati, M. Pascher et al.}

\begin{abstract}

Despite the growth of physically assistive robotics (PAR) research over the last decade, nearly half of PAR user studies do not involve participants with the target disabilities. There are several reasons for this---recruitment challenges, small sample sizes, and transportation logistics---all influenced by systemic barriers that people with disabilities face. However, it is well-established that working with end-users results in technology that better addresses their needs and integrates with their lived circumstances.  In this paper, we reflect on multiple approaches we have taken to working with people with motor impairments across the design, development, and evaluation of three PAR projects: (a) assistive feeding with a robot arm; (b) assistive teleoperation with a mobile manipulator; and (c) shared control with a robot arm. We discuss these approaches to working with users along three dimensions---individual- vs. community-level insight, logistic burden on end-users vs. researchers, and benefit to researchers vs. community---and share recommendations for how other PAR researchers can incorporate users into their work.

\end{abstract}




\maketitle
\renewcommand{\thefootnote}{\fnsymbol{footnote}}
\footnotetext[1]{Both authors contributed equally to this paper}
\renewcommand{\thefootnote}{\arabic{footnote}}

\begin{figure*}[t!]
    \centering
    \includegraphics[width=\textwidth]{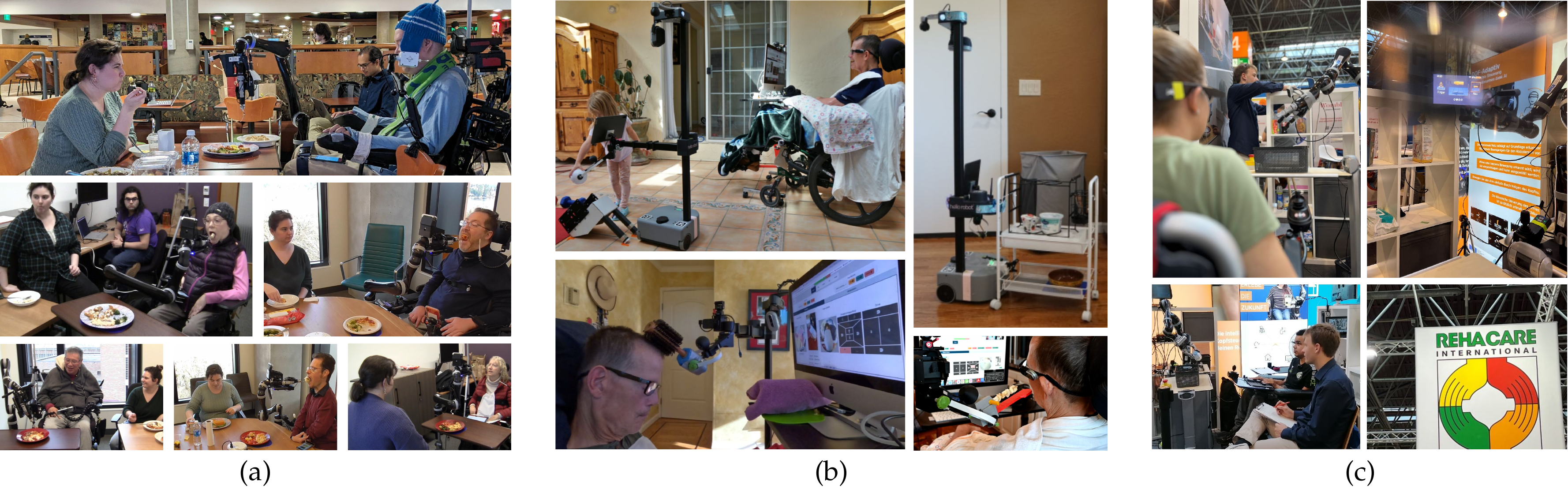}
    \caption{The three projects highlighted in this work: (a) assistive feeding with a robot arm, (b) assistive teleoperation with a mobile manipulator, (c) shared control with a robot arm.}
    \label{fig:projectsOverview}
\end{figure*}

\section{Introduction}
When designing, developing, and evaluating \acp{PAR} for people with motor impairments, it is crucial to work with them to ensure the technology addresses their needs and integrates with their lived circumstances~\citep{mankoff2010disability,Pascher.2021recommendations}. However, only about half of \ac{PAR} user studies, and less than half of summative studies\footnote{Summative studies evaluate, or ``sum up,'' a technology, whereas formative studies generate early insights that ``form'' the design of a technology.}, 
incorporate members of the target community~\citep{nanavati2023physically}. Reasons for this include: 
(1) outreach and recruitment of people with disabilities is challenging because they tend to have lower technology access and usage~\citep{pew2021DisabilityTechnologyAccess}; (2) systemic barriers hinder people with disabilities' access to college education~\citep{bls2015DisabilityEducationLevel}, making them underrepresented amongst a population that is commonly used for studies; and (3) 
coordinating travel to the research venue is challenging, as transportation access and usage tends to be lower amongst people with disabilities~\citep{dot2022DisabilityTransportationAccess}. These factors extend the time it takes to do rigorous studies with participants with motor impairments, which can be incompatible with rapid research timelines.



\section{Contribution}
We critically reflect on our experiences and methodological insights working with users with motor impairments 
across three \ac{PAR} projects: \textbf{(1)} assistive feeding with a robot arm; \textbf{(2)} assistive teleoperation with a mobile manipulator; and \textbf{(3)} shared-control with a robot arm. These projects were conducted independently by 3 different research labs, and the insights emerged from joint discussions. Across the projects, we employed diverse ways of engaging end-users: community research, remote studies, home deployments, and an in-the-wild study at a trade fair. 
We discuss these approaches along three key dimensions: 

\begin{enumerate}
    \item \textbf{Individual- vs. Community-Level Insights}: 
    \textit{How should we balance between conducting deep research with few participants versus broad research with many participants?}
    \item \textbf{Logistical Burdens on End-Users vs. Researchers}: \textit{How can we navigate the differential logistical burdens that end-users and researchers face during a user study?}
    \item \textbf{Benefits to Researchers vs. Community}: 
    \textit{What benefits do researchers and end-users get from study participation? How should that knowledge influence researchers' work with end-users?}
\end{enumerate}
\section{Related Work}
In recent years, there has been growing recognition of the importance of user participation in the design of assistive technologies.
\citet{Thielke.2012} and \citet{Merkel.2018} expressed the need for this collaborative approach to maximize user acceptance, and presented various methods for integrating users, family members, caregivers, and assistants into the innovation process.

\subsection{Participatory Design}
Assistive technologies are becoming a vital factor in assisted living, minimizing the need for constant caregiver presence and restoring some independence to people with motor impairments~\cite{Park.2020,Kyrarini.2021}.
However,  assistive technologies 
can have high rates of non-acceptance and non-use~\citep{Scherer.2005, Verza.2006}, partly due to end-user exclusion during the design process~\citep{Federici.2015}. 
Participatory Design---where developers work collaboratively with end-users (and/or their advocates) throughout the design process~\citep{Lee.2017, Simonsen.2013}---has become a recommended way to increase acceptance of the final product~\cite{Vines.2013}. 
All three projects in this paper involve such a collaborative approach. 
Across all projects, we found that participants valued their inclusion in the design process of a device developed specifically for them.


\subsection{Empowerment Design}
In the words of \citet{ladner2015design}, ``some of the best work comes when there are people with disabilities on the design and development team, contributing to all aspects of the design and implementation...I call this strong engagement by users \emph{design for user empowerment}, meaning, in its strongest sense, that the users of the technology are empowered to solve their own accessibility problems.''

Empowering users with disabilities in \ac{PAR} research is crucial\footnote{This is exemplified in the disability rights slogan ``nothing about us without us''~\citep{charlton1998nothing}}, so they are not just brought in as part of a few studies, but are instead empowered to design and build the technologies themselves~\cite{ertner2010five}. Empowerment Design  focuses on developing socio-technical structures to provide that empowerment
~\cite{barab2002empowerment}. Several methods discussed in this work, although not originally conceptualized as empowerment design, include empowerment as a core focus.

\section{Descriptions of the Three Projects}
Three projects (\autoref{fig:projectsOverview}) 
underlie this paper's critical reflection. 

\subsection{Assistive Feeding with a Robot Arm}

The ultimate goal of this project is to develop a robot-assisted feeding system that users with motor impairments can use to feed themselves an entire meal of their choice, in an environment of their choice, without researcher interventions, while aligning with their preferences. The system consists of a \emph{Kinova JACO Gen2}~\cite{kinova} robot arm with an RGB-D eye-in-hand camera. The robot's gripper holds a custom 3D-printed fork with an attached force-torque sensor. The system is fully portable; it can be mounted onto and powered from standard power wheelchairs. This project has been ongoing 
for ~10 years, with 100s of user-hours in co-designing and evaluation. 

The most recent iteration of system design, development, and evaluation---over the last two years---involved end-users in two main capacities. The first is as a community researcher; we worked with two end-users throughout the entire design and research process, from ideation to dissemination, following principles from \ac{CBPR}\footnote{In \ac{CBPR}, academics work equitably with community members throughout all research stages. 
\ac{CBPR} is used in health~\cite{israel1998review, viswanathan2004community} and assistive technology~\cite{nanavati2023design, chapko2020we, kushalnagar2020teleconference, bell2019collaborative} research.\label{foot:cbpr}}~\citep{nanavati2023design, nanavati2024lessons}. 
The second is as traditional participants; we invited $6$ people with motor impairments to come to out-of-lab environments (a cafeteria, conference room, and office) to eat a meal with us, using the robot~\citep{nanavati2024lessons}. 

\subsection{Assistive Teleoperation with a Mobile Manipulator}

The ultimate goal of this project is to develop an assistive teleoperation interface that empowers people with motor impairments to complete their desired tasks around the house. These tasks include: physically assisting oneself (e.g., itch-scratching, self-feeding); 
physically assisting one's caregiver (e.g., transporting a meal or laundry 
cart to/from the bedroom); and participating in social interactions (e.g., playing poker, playing indoor basketball with a child)~\citep{ranganeni2024robots}. The system consists of a \emph{Stretch} mobile manipulator~\citep{kemp2022design} and a cursor-based, customizable teleoperation interface~\citep{ranganeni2023evaluating}. 

The design, development, and evaluation of the teleoperation interface involved end-users in two main capacities. First, $10$ users with motor impairments participated in a remote study; they were not co-located with the robot but used the teleoperation interface to have the robot perform kitchen tasks
~\citep{ranganeni2023evaluating}. Second, we conducted multiple week-long deployments and co-design sessions in the home of one end-user. 
We identified tasks he wanted to do, extended the system so the robot could complete those tasks, allowed him to freely use the system, and iteratively incorporated his feedback.

\vspace{-1mm}
\subsection{Shared-Control with a Robot Arm}
This project began with an ethnographic study involving semi-structured interviews and in-situ observations in participants' homes to understand 
how an assistive robotic arm could support them. 
The analysis revealed: (1) participants' desire for alone-time without caregiver assistance; (2) their wariness about fully autonomous robots' failures; (3) their desire to always be in control of robotics aids; and (4) 
most participants only have access to 1 or 2 \acp{DoF} to control a robot arm. 
Thus, we developed a shared-control approach~\cite{Pascher.2024adaptix} to allow users to interact with a robotic arm for everyday tasks (e.g., picking up an object, opening a door)~\citep{Pascher.2023inTimeAndSpace}. The approach utilizes a convolutional neural network to perceive the visual scene and suggest input mappings, thereby reducing the control complexity of a 7-\acp{DoF} robot arm to 2 input-\acp{DoF}. 
The \ac{DoF} mapping suggestions are visualized in an arrow-based representation so users can understand the robot's motion intent~\citep{Pascher.2023robotmotionintent}.
The user interacts with the robot through head motion and receives visual feedback 
through a \emph{Google Glass} interface. 
This project was heavily informed by a ``lead user'' with motor impairments we met regularly to discuss the research direction and system design.


To evaluate the final prototype of this system, we sought access to many participants with diverse motor impairments. Thus, we 
rented a booth at the \emph{REHACARE},\footnote{\emph{REHACARE} trade fair. \url{https://www.rehacare.de}, last retrieved \today.} trade fair
, one of the largest trade fairs in rehabilitation and care. 
This allowed us to conduct individual study sessions with 24 participants, with a diverse variety of motor impairments or related disabilities, over just four days. 
\vspace{-1mm}
\section{Discussion}
Here we reflect upon our research methods along three dimensions.

\vspace{-1mm}
\subsection{Individual- vs. Community-Level Insights}

Given the reality of time and resources constraints, there is often a trade-off between conducting deep research that involves long-term engagement with a few end-users, versus broad research that involves shorter-term engagements across many end-users.

Conducting \emph{deep research with a few individuals} can result in insights on the nuances of their preferences and environments, which is necessary to develop robots that work very well for those specific people (which is arguably the goal of personal robotics). However, such research requires a large commitment of time and relationship-building per individual. 

Conducting \emph{broad research across many individuals} can result in insights on the diversity of needs and wants in the community, which is necessary to develop a robot that can be distributed at scale. 
However, this involves spending just a few hours per participant, not long enough to understand individual-level nuances. 


\subsubsection{Methods}\label{sec:individualVsCommunityHowWeNavigated}

Across all projects, we navigated this dimension by working deeply with few users \emph{for the full duration of the project}, and working broadly across several users \emph{for specific studies}. 

We found multi-faceted benefits of working deeply with a few individuals for the entirety of the project:
\begin{itemize}
    \item They taught us nuances of their lived experiences that we may not have expected: e.g. day-to-day vocal strength varies, influencing the accuracy of speech assistive technologies. Understanding these nuances led to a more user-centered design.
    \item They grew familiar with the technology, providing glimpses into end-user perspectives after the novelty effect wears away.
    \item Our partnership lowered the barrier of entry for new students to do user-centered work, since the relationship was established.
\end{itemize}
How can researchers find long-term partners from the community? Across all projects, these individuals were technology enthusiasts. In one case, they were participants in past studies who wanted to get more involved. In the other cases, they heard about the research (e.g., through media) and reached out to get more involved.

To complement the deep research with few individuals, at multiple points within the projects (both for formative and summative studies) we conducted broad research across many community members. We recruited participants with motor impairments by: (a) contacting local organizations that work with them; (b) posting on relevant social media groups; (c) traveling to events frequented by target users (e.g., an assistive technology trade fair); and (d) word of mouth from other participants. Community researchers also helped by connecting us to participants and even co-running a study~\citep{nanavati2023design}.


\subsubsection{Key Takeaways}
\begin{enumerate}
    \item A research project is a long time, often spanning several studies. There is room to \emph{combine both deep and broad research}.
    \item Technology enthusiasts from the target community  may be excited to work deeply with you over the long-term. One way of finding such individuals is presenting the project at events that are geared towards the target community.
    \item Recruitment for deep and broad research is complementary; broad research can connect you with people to work deeply with, and they in turn can connect you to future participants. 
\end{enumerate}



\subsection{Logistical Burdens on Users vs. Researchers}

Research with end-users inevitably involves logistical burdens. On the one hand, asking participants to come in-person to a lab where the robot already is places considerable logistical burden on the participant. They may need to schedule a caregiver, arrange a vehicle, account for transportation delays, shift their daily schedule (e.g., morning routines to get ready, mealtimes, medicine-taking times) to accommodate the study, and more. Although research labs often compensate participants for their time spent in the study, and sometimes for transportation costs to get to/from the  lab, that does not account for the time and mental energy spent coordinating the aforementioned logistics. However, a research benefit of this approach is that it is easier to run controlled experiments in a lab, and robots typically work better in their development environment. 

On the other hand, bringing the robot to the end-user shifts the logistical burden onto the researchers. It is challenging to develop a robot that is portable enough to be transported, robust enough to work reliably in a new location, and flexible enough to be debugged on the fly. 
Transporting and setting up the robot in a new environment (e.g., power supply, networking, mapping the environment, etc.) is challenging. However, benefits of this approach are that it is easier for participants and has more ecological validity. 

\subsubsection{Methods}

The assistive feeding project navigated this dimension by conducting formative research remotely over video call~\citep{nanavati2023design}, so the only logistical burden on participants was getting set up on their computer at home. 
In our most recent in-person study~\citep{nanavati2024lessons}, we had participants travel to our campus, but held the study in an out-of-lab venue (e.g., cafeteria, conference room, office). Although this still places a high logistical burden on participants, it can be a stepping stone to home deployments. 
We designed the system to make it easier to transport and set up out-of-lab, by making it portable with an easy customization interface.

The assistive teleoperation project has conducted multiple home deployments~\citep{ranganeni2024robots}. This shifts much of the burden to the researcher; the chief logistical burden for the participant and their caregiver(s) is hosting researchers. We also ran studies where participants joined remotely and teleoperated the robot~\citep{ranganeni2023evaluating}. This approach had lower logistical burden on both the researcher and user, and geographically expanded the pool of potential participants. 

The shared control project found another way of reducing the logistical burden on participants; we went to an event that is well-attended by members of the target community. This provided access to many participants, and enabled us to run a study with high ecological validity. However, in practice, this approach came with additional challenges on the researchers. 
The dynamic environment 
brought difficulties with wireless communication, light pollution, noise, and interruptions. Each participant required a slightly different technology setup. These challenges required adaptation on-the-fly, 
and we could not reschedule participants; any robot down-time directly reduced the number of study participants.

\subsubsection{Key Takeaways}
\begin{enumerate}
    \item Remote studies can decrease the logistical burden on participants without increasing the logistical burden on researchers.
    \item When studies require co-location (i.e., cannot be done remotely):
    \begin{enumerate}
        \item If many participants are desired, go to a venue the community frequents, like an assistive technology trade fair.
        \item If fewer participants are acceptable, do an in-lab study if you desire more internal validity, or an out-of-lab study or home deployment if you desire more ecological validity.
    \end{enumerate}
    \item When bringing the robot to another location, take preemptive system design steps to make it easier to transport, setup, and use the robot in a new and dynamic environment.
\end{enumerate}

\subsection{Benefits to Researchers vs. Community}

The fact that researchers get benefits from research is self-evident; it can lead to career growth, access to funding, fame and renown, future research opportunities, and more. End-users are no doubt indispensable to \ac{PAR} research---they influence our system design, system evaluations, and the key insights that motivate future work. However, what benefits do end-users get from their participation, how does that align with the benefits they would like to get, 
and is that commensurate with their indispensable role in \ac{PAR} research?

Compensation is one benefit for participants; however, as mentioned above, it may not cover the additional logistical burdens involved in participating in a research study. Some people participate to learn about robotics. Others because they want to accelerate research that benefits their community. Others may not have an explicitly stated motivation, but the user study might unearth ways the researchers can support them beyond the study. 

\subsubsection{Methods}

In the assistive feeding work, we sought ways to benefit participants beyond just using the robot. One participant was interested in learning technical details of how the robot acquired food---we explained this as part of the social conversation during her meal. For another participant, we showed him how to use an assistive technology feature he did not know about. 
We also sought opportunities to cross-pollinate across participants; one participant had open-sourced a 3D-printed tool for self-feeding, which we shared with other participants with similar mobility. 

The idea of benefiting end-users beyond the research was particularly salient with a community researcher. One time, he was struggling with manually populating a mailing list from a spreadsheet; we wrote a script to automate that. Another time, he was appealing denied health insurance coverage;  we connected him with resources to help. He also mentioned additional benefits, such as stronger grant applications after he co-authored a paper with us. 

In the assistive teleoperation work, we developed our in-home deployment to match the needs and expectations of the user. For example, the user expressed a desire to play with his granddaughter, since he had never played with her before and she barely acknowledged him. After three play sessions with his granddaughter using the robot (\autoref{fig:projectsOverview}b) he has developed a relationship with her. 

In the shared control work, it was important to accept the diverse reasons that participants came to our booth; although some were interested in the study, others just wanted to talk, which required researchers to subordinate the study goals for the broader goal of sharing knowledge with the community. 
Further, since the trade fair had commercial products, it was important to manage expectations by clarifying that this robot may not be available anytime soon. 

Across all three projects, having a dedicated focus on benefiting participants beyond the study helped to strengthen ties, build empathy, and gain broader exposure within the community, all of which strengthened long-term project goals.



\subsubsection{Key Takeaways}
\begin{enumerate}
    \item Be attuned to ex/implicit signs of how you can benefit participants beyond the study; where possible, provide that benefit.
    \item Working with the user to co-design the robot's task can increase the likelihood of them benefiting from the study in the present.
    \item Focusing on helping the community more broadly 
    has long-term positive impacts on a project, due to a positive relationship between researchers and members of the target community.
\end{enumerate}

\subsection{Limitations and Future Work}

This work presents key dimensions for working with end users in \ac{PAR} research.
A limitation is that the three projects were conducted independently and the synthesis was done afterward; thus, the  study do not follow a cohesive research agenda
An exciting direction for future work is expanding these dimensions into a framework
to clarify how researchers can situate themselves at different stages when conducting this type of research. 

\begin{acks}
This research is (partially) funded by the National Science Foundation GRFP (DGE-1762114), NRI (\#2132848) and CHS (\#2007011), German Federal Ministry of Education and Research (BMBF, FKZ: \href{https://foerderportal.bund.de/foekat/jsp/SucheAction.do?actionMode=view&fkz=16SV8565}{16SV8565}, \href{https://foerderportal.bund.de/foekat/jsp/SucheAction.do?actionMode=view&fkz=16SV7866K}{16SV7866K}, \href{https://foerderportal.bund.de/foekat/jsp/SucheAction.do?actionMode=view&fkz=13FH011IX6}{13FH011IX6}), the Office of Naval Research (\#N00014-17-1-2617-P00004, \#2022-016-01 UW), and Amazon.
\end{acks}
\newpage
\bibliographystyle{ACM-Reference-Format}
\balance
\bibliography{MainPaper}

\end{document}